\documentclass
[superscriptaddress,secnumarabic,amssymb,amsmath,nobibnotes,aps,prd,showkeys,showpacs,nofootinbib,onecolumn,eqsecnum]{revtex4}%
\usepackage{graphics}
\usepackage{graphicx}
\usepackage{color}
\usepackage{epsf}
\usepackage{bm}
\usepackage{amsmath,amssymb,amsfonts,mathrsfs,amsthm}
\usepackage{latexsym}
\usepackage{enumerate}
\usepackage{comment}
\usepackage{hyperref}
\usepackage{amsmath}
\usepackage{amsfonts}
\usepackage{amssymb}
\usepackage{epstopdf}
\usepackage{setspace}%
\providecommand{\U}[1]{\protect\rule{.1in}{.1in}}

\newcommand{\be}{\begin{equation}}
\newcommand{\ee}{\end{equation}}
\newcommand{\bea}{\begin{eqnarray}}
\newcommand{\ea}{\end{eqnarray}}
\newcommand{\ben}{\begin{equation*}}
\newcommand{\een}{\end{equation*}}
\newcommand{\bean}{\begin{eqnarray*}}
\newcommand{\eean}{\end{eqnarray*}}
\def\bal#1\eal{\begin{align}#1\end{align}}

\newcommand{\mincir}{\raise
-3.truept\hbox{\rlap{\hbox{$\sim$}}\raise4.truept\hbox{$<$}\ }}
\newcommand{\magcir}{\raise
-3.truept\hbox{\rlap{\hbox{$\sim$}}\raise4.truept\hbox{$>$}\ }}

\begin{document}
\title{Cosmological Solutions in Multiscalar Field Theory}
\author{N. Dimakis}
\email{nsdimakis@scu.edu.cn}
\email{[]nsdimakis@gmail.com}
\affiliation{Center for Theoretical Physics, College of Physical Science and Technology
Sichuan University, Chengdu 610065, China}
\author{A. Paliathanasis}
\email{anpaliat@phys.uoa.gr}
\affiliation{Institute of Systems Science, Durban University of Technology, POB 1334 Durban
4000, South Africa.}
\author{Petros A. Terzis}
\email{pterzis@phys.uoa.gr}
\affiliation{Nuclear and Particle Physics section, Physics Department, University of
Athens, 15771 Athens, Greece}
\author{T. Christodoulakis}
\email{tchris@phys.uoa.gr}
\affiliation{Nuclear and Particle Physics section, Physics Department, University of
Athens, 15771 Athens, Greece}

\begin{abstract}
We consider a cosmological model with two scalar fields minimally coupled to
gravity which have a mixed kinetic term. Hence, Chiral cosmology is included
in our analysis. The coupling function and the potential function, which
depend on one of the fields, characterize the model we study. We prove the
existence of exact solutions that are of special interest for the cosmological
evolution. Furthermore, we provide with a methodology that relates the
scale factor behaviour to the free functions characterizing the scalar field
kinetic term coupling and potential. We derive the necessary conditions that
connect these two functions so that the relative cosmological solutions can be
admitted. We find that unified dark matter and dark energy solutions are
allowed by the theory in various scenarios involving the aforementioned functions.

\end{abstract}
\maketitle
\date{\today}

\section{Introduction}

From the detailed analysis of the recent cosmological data
\cite{Eisenstein:2005su,Bennett:2012zja,Ade:2015xua} it has been made clear
the necessity for the introduction of an exotic fluid term with negative
pressure in Einstein's General Relativity. The role of this exotic matter
source is to explain the acceleration phases of the universe. So far its
nature is unknown and it has only been observed indirectly; it is thus
referred in the literature as dark energy. The contribution of the dark energy
in the observable universe is approximately $\sim70\%$ while the rest
$\sim30\%$ corresponds mostly to dark matter (with a smaller contribution by
baryonic matter and radiation). Dark matter is a pressureless fluid source
(dust) that also has not been observed directly and which was introduced in
order to explain the observed rotation curves of the galaxies.

Although the cosmological constant is the most simple dark energy candidate,
$\Lambda$-cosmology suffers by two major drawbacks: a) fine tuning and b) the
coincidence problem \cite{Weinberg89,Peebles03}. A quite wide-known
alternative approach on the cosmological constant is scalar field cosmology
\cite{Ratra88}. Scalar fields can be introduced either minimally or
nonminimally coupled to gravity. In what regards the minimally coupled case,
the general solution for an arbitrary potential function has been obtained for
all types of Friedmann--Lema\^{\i}tre--Robertson--Walker (FLRW) spacetimes:
open, closed and spatially flat \cite{genFLRW}. Scalar fields can also be used
to incorporate - in an equivalent picture - the new degrees of freedom
provided by the higher-order derivatives in extended theories of gravity
\cite{bran01,ohanlon,hor,gali,bar01,lan001}. Apart from the application of
scalar fields in order to account for the late-time acceleration of the
universe, they have been additionally used as unified dark energy models
\cite{ref1,newinf,ref4,peebles,tsuwi,vs,urena,ber01,ber02,ber03,ber04}.

With the term scalar field cosmology usually someone refers to quintessence
theories. However, there are many alternatives like, phantom fields, Hordenski
gravity \cite{hor}, or even multi-scalar field models. The latter have a broad
range of applications as they can provide alternative models of inflation,
namely: hybrid inflation, $\alpha-$attractors, double inflation and many
others \cite{hy1,hy2,hy3,hy4,atr1,atr2,atr4,atr5,atr6}. Quintom scalar field
models consist of two scalar fields, a quintessence field and a phantom field,
for a review we refer the reader in \cite{quin00}. The main characteristic of
the quintom models is that in general the two scalar fields are minimally
coupled and only a few models have been proposed where an interaction term
between the two fields is introduced in the potential \cite{quin00}. While a
quintom model with a mixed-kinetic term has been introduced before
\cite{qui00}, due to the fact that the geometry of the space where the kinetic
terms of the scalar field form is flat there can always be defined new scalar
fields so as to neglect the mixed kinetic term.

Multi-scalar field models where interaction exists in the kinetic term of the
Lagrangian is where this work focuses. A specific family of these models,
where the two scalar fields defined into a two-dimensional space of constant
nonvanishing curvature, are the Chiral cosmological models
\cite{atr6,chir1,chir3} which are directly related with the non-linear sigma
cosmological model \cite{sigm0a,sigm0,sigm1}. In this work we consider
two-scalar field cosmological model where the one field is quintessence and
the second field interact with the quintessence only in the kinetic term. The
model consists of two unknown functions, the potential which drives the
dynamics of the quintessence and the interaction between the fields. For this
cosmological model we prove the existence of particular solutions of great
interest and determine the conditions that they impose in the functions
characterizing the theory. Among the solutions we study are: scaling
solutions, de Sitter solutions and that of asymptotic de Sitter spacetimes
with or without cosmological singularity. The last cosmological scenario is
very interesting since $\Lambda$CDM cosmology is recovered. What is more, the
determination of exact or analytical solutions is important because the main
properties of a given model can be determined without suffering from the
problem of selecting the initial conditions. Moreover, when we use numerical
methods to solve a given dynamical system, we know that the numerical
simulations describe real trajectories of the problem. Some results with exact
and analytic solutions in multi scalar field cosmology are given in
\cite{chir1,chir3,no1,kess,ss1,mf1,mf2,mf3,mf4,mf5,mf6,mf7,mf8,mf9,mf10,mf11}.

In this work we study a two scalar field cosmological system. We are
mainly interested in presenting a method for providing exact solutions
together with the class of theories to which they correspond. This is realized
by deciding on a specific scale factor behaviour and then providing
algorithmically the relation between the theory's free functions that permit
such a solution. Thus, we are able to derive - within the type of model we
study - which specific relation between the kinetic coupling and the scalar
field potential allows for certain interesting cosmological solutions.

The outline of the paper is as follows: In Section \ref{sec2} we briefly
present the cosmological model that we consider and derive the gravitational
field equations in the case of a spatially flat FLRW spacetime. In Section
\ref{sec3} we prove the existence, under conditions, of four important
families of scale factor solution of cosmological interest. More specifically,
we determine solutions for scaling solutions, de Sitter solution,
asymptotically de Sitter spacetimes and exponentials of arbitrary powers of
time. There exists a case where the scalar field potential and the interaction
function are exponential with different exponents, where the scalar factor is
given by a power-law function in which the acceleration of the universe can be
described, including the inflationary era. For that specific cosmological
model in Section \ref{sec4} we perform a detailed analysis on the dynamics of
field equations by studying the existence and the stability of critical
points. Finally, in Section \ref{sec5} we discuss our results and draw our conclusions.

\section{Field equations}

\label{sec2}

Let us consider the action integral
\begin{equation}
S=\int\!\!d^{4}x\sqrt{-g}\mathcal{L}\left(  R\left(  x^{\mu}\right)
,\phi\left(  x^{\mu}\right)  ,\psi\left(  x^{\mu}\right)  ,\nabla_{\nu}%
\phi\left(  x^{\mu}\right)  ,\nabla_{\nu}\psi\left(  x^{\mu}\right)  \right)
\label{action}%
\end{equation}
characterized by the Lagrangian density
\begin{equation}
\mathcal{L}=\frac{1}{2}R-\frac{1}{2}\nabla_{\mu}\phi\nabla^{\mu}\phi
-\frac{F(\phi)}{2}\nabla_{\mu}\psi\nabla^{\mu}\psi-V(\phi), \label{Lagden}%
\end{equation}
where we have adopted the geometric units $8\pi G=c=1$. As usual, $g$ is used
to symbolize the determinant of the space-time metric, while $R$ is reserved
for the scalar curvature.

From \eqref{Lagden} it is obvious that we consider two scalar fields
$\phi\left(  x^{\mu}\right)  $ and $\psi\left(  x^{\mu}\right)  $ as the
matter content of the theory. At the same time we assume an interacting
function $F\left(  \phi\left(  x^{k}\right)  \right)  $ and a potential
$V\left(  \phi\left(  x^{\mu}\right)  \right)  $ that depend solely on the
scalar field $\phi\left(  x^{\mu}\right)  $.

The field equations produced in this setting, by variation with respect to the
metric $g_{\mu\nu}\left(  x^{\mu}\right)  $ and the scalars $\phi\left(
x^{\mu}\right)  $, $\psi\left(  x^{\mu}\right)  $ are respectively:
\begin{subequations}
\label{eqofmopsi}%
\begin{align}
R_{\mu\nu}-\frac{1}{2}g_{\mu\nu}R  &  =T_{\mu\nu}\label{eqofmoein}\\
\nabla_{\mu}\nabla^{\mu}\phi-V^{\prime}(\phi)-\frac{1}{2}F^{\prime}%
(\phi)\nabla_{\mu}\psi\nabla^{\mu}\psi &  =0\\
\nabla_{\mu}\left(  F(\phi)\nabla^{\mu}\psi\right)   &  =0,
\end{align}
where $R_{\mu\nu}$ is the Ricci tensor and the prime denotes a derivative with
respect to $\phi$. The energy momentum tensor that appears in
\eqref{eqofmoein} is given by
\end{subequations}
\begin{equation}
T_{\mu\nu}=\nabla_{\mu}\phi\nabla_{\nu}\phi-\frac{1}{2}\left(  \nabla^{\kappa
}\phi\nabla_{\kappa}\phi\right)  g_{\mu\nu}-V(\phi)+F(\phi)\nabla_{\mu}%
\psi\nabla_{\nu}\psi-\frac{F(\phi)}{2}\left(  \nabla^{\kappa}\psi
\nabla_{\kappa}\psi\right)  g_{\mu\nu}. \label{enmomt}%
\end{equation}

Due to the fact that we choose the functions $F\left(  \phi\left(  x^{\mu
}\right)  \right)  $, $V\left(  \phi\left(  x^{\mu}\right)  \right)  $ to
depend only on $\phi\left(  x^{\mu}\right)  $, the action \eqref{action} is
bound to remain form invariant under the global transformation $\psi\left(
x^{\mu}\right)  \mapsto\psi\left(  x^{\mu}\right)  +\varepsilon$, in which
$\varepsilon$ is a constant. The corresponding conservation law is already
obvious from the equation of motion \eqref{eqofmopsi} which implies
\begin{equation}
\partial_{\mu}\left[  \sqrt{-g}F(\phi)\nabla^{\mu}\psi\right]  =0,
\end{equation}
where the term inside the bracket is the Noether current corresponding to the
aforementioned symmetry transformation.

We consider a spatially flat FLRW spacetime with line element
\begin{equation}
ds^{2}= g_{\mu\nu}dx^{\mu}dx^{\nu}=-N^{2}(t)dt^{2}+a(t)^{2}\left(
dr^{2}+r^{2}d\theta^{2}+r^{2}\sin^{2}\theta d\varphi^{2}\right)  ,
\end{equation}
where $a\left(  t\right)  $ is the scale factor of the universe and $N\left(
t\right)  $ is the lapse function. For the scalar fields we assume that they
inherit the symmetries of the spacetime which means $\phi=\phi(t)$ and
$\psi=\psi(t)$. Under this assumption the gravitational plus matter field
equations reduce to the following set of ordinary differential equations
\begin{subequations}
\label{odes}%
\begin{align}
\frac{1}{2N^{2}}\left[  F(\phi)\dot{\psi}^{2}+\dot{\phi}^{2}-\frac{6\dot
{a}^{2}}{a^{2}}\right]  +V(\phi)  &  =0\label{con}\\
\frac{1}{2N^{2}}\left[  -\frac{4\ddot{a}}{a}+\frac{4\dot{a}\dot{N}}{aN}%
-\frac{2\dot{a}^{2}}{a^{2}}-F(\phi)\dot{\psi}^{2}-\dot{\phi}^{2}\right]
+V(\phi)  &  =0\label{ode1}\\
\frac{1}{2N^{2}}\left[  2\ddot{\phi}+\frac{6\dot{a}\dot{\phi}}{a}-F^{\prime
}(\phi)\dot{\psi}^{2}-\frac{2\dot{N}\dot{\phi}}{N}\right]  +V^{\prime}(\phi)
&  =0\label{ode2}\\
\frac{d}{dt}\left(  \frac{a^{3}F(\phi)\dot{\psi}}{N}\right)   &  =0,
\label{ode3}%
\end{align}
where the dot signifies derivative with respect to the parameter $t$.

From the last equation the corresponding conserved quantity of the reduced
system becomes apparent since it implies
\end{subequations}
\begin{equation}
Q:=\frac{a^{3}F(\phi)\dot{\psi}}{N}=c_{1}, \label{Qconst}%
\end{equation}
with $c_{1}$ being a constant of integration.

It is interesting to note that the same situation can be reproduced in the
context of a minisuperspace approximation through the Lagrangian function
\begin{equation}
L=-\frac{1}{2N}\left(  6a\dot{a}^{2}-a^{3}\dot{\phi}^{2}-a^{3}F(\phi)\dot
{\psi}^{2}\right)  -Na^{3}V(\phi), \label{LagN}%
\end{equation}
whose Euler-Lagrange equations are equivalent to \eqref{odes} and of course
its symmetries lead to the same conserved charge $Q=\frac{\partial L}%
{\partial\dot{\psi}}$.

We now proceed with our analysis by investigating the conditions under which
certain interesting gravitational solutions may emerge. In what follows we
adopt the gauge choice $N=1$ so that the results are expressed directly in a
cosmological time.

\section{Exact solutions}

\label{sec3}

The gravitational field equations form a constraint system of nonlinear
second-order differential equations for the three dependent variables
$\left\{  a\left(  t\right)  ,\phi\left(  t\right)  ,\psi\left(  t\right)
\right\}  $. There are many alternative ways to study the dynamics and the
evolution of the field equations. In this work we are interested in the
existence of exact solutions. More specifically, we assume that the physical
space is specified by particular functional forms for the scalar factor
$a\left(  t\right)  $ which describe various eras in the evolution of the
universe. Such an analysis is important because it provides information for
the existence of various cosmological eras for the specific cosmological
model. In addition to this, the cosmological validity of each model can be
studied directly from such an analysis.

From the conserved quantity \eqref{Qconst} we easily deduce that
\begin{equation}
\dot{\psi}=\frac{c_{1}}{a^{3}F(\phi)}. \label{psidot}%
\end{equation}
By adding \eqref{con} and \eqref{ode1} the relation
\begin{equation}
V(\phi)=\frac{\ddot{a}}{a}+\frac{2\dot{a}^{2}}{a^{2}} \label{eqV}%
\end{equation}
is obtained (recall that we have set $N=1$). At this point we have acquired
the potential $V(\phi)$ as a pure function of $t$. We continue our analysis by
considering several interesting situations regarding the scale factor.

\subsection{Power law solutions}

In this case we are interesting in studying the conditions that a solution of
the form
\begin{equation}
a(t)=t^{\sigma},\quad\sigma\in\mathbb{R}, \label{apower}%
\end{equation}
can set over the theory. The power-law solution (\ref{apower}) is important,
because it can describe the matter dominated era solution~$\left(
\sigma=\frac{2}{3}\right)  $, the radiation era solution~$\left(  \sigma
=\frac{1}{2}\right)  $, or the early accelerated phase of the universe when
$\sigma>1$.

For this choice of scale factor the space-time exhibits a curvature
singularity at $t=0$ for any value of $\sigma$. With the help of
\eqref{apower} relation \eqref{eqV} results in
\begin{equation}
V(\phi)=\frac{\sigma(3\sigma-1)}{t^{2}}. \label{Vtpower}%
\end{equation}
This leads us to consider two distinct situations depending on the value of
$\sigma$.

\subsubsection{Case
\texorpdfstring{$\sigma\neq\frac{1}{3}$}{sigma not equal to 1 over 3}}

We start by considering that the potential $V(\phi)$ cannot be zero, which
according to \eqref{Vtpower} corresponds to $\sigma\neq\frac{1}{3}$. By
inverting equation \eqref{Vtpower} we obtain $t$ as a function of $\phi$,
i.e.
\begin{equation}
t=\pm\frac{\sqrt{\sigma}\sqrt{3\sigma-1}}{\sqrt{V(\phi)}}. \label{tVpower}%
\end{equation}
If we additionally take the derivative of \eqref{Vtpower} with respect to
time, the relation
\begin{equation}
V^{\prime}(\phi)\dot{\phi}=\frac{2(1-3\sigma)\sigma}{t^{3}}\Rightarrow
\dot{\phi}=\frac{2\sigma(1-3\sigma)}{t^{3}V^{\prime}(\phi)}, \label{phiV1}%
\end{equation}
emerges. Direct substitution of \eqref{psidot}, \eqref{tVpower} and
\eqref{phiV1} into the constraint equation \eqref{con} leads to the
expression
\begin{equation}
F(\phi)=c_{1}^{2}\sigma^{1-3\sigma}(3\sigma-1)^{1-3\sigma}\frac{V(\phi
)^{3\sigma-1}V^{\prime}(\phi)^{2}}{2(\sigma V^{\prime}(\phi)^{2}-2V(\phi
)^{2})}, \label{FVpower}%
\end{equation}
which involves only the two functions of $\phi$. This is how $F\left(
\phi\right)  $ and $V\left(  \phi\right)  $ have to be related for a power law
solution of the form \eqref{apower} with $\sigma\neq\frac{1}{3}$ to be
admitted by the system. Any choice for the potential $V(\phi)$ leads to the
corresponding expression of the coupling $F(\phi)$ needed for such a solution.
Let us proceed with the study of two particular examples.

\vspace{3pt}

\paragraph{Example 1: Power law potential.}

Let us set $V(\phi)=V_{0}\phi^{\kappa}$, with $V_{0}$ and $\kappa$ constants.
Then, \eqref{FVpower} implies
\begin{equation}
F(\phi)=F_{0}\frac{\phi^{\kappa(3\sigma-1)}}{2(\kappa^{2}\sigma-2\phi^{2})},
\label{FVpower1}%
\end{equation}
where with $F_{0}$ we re-parametrized the constant of integration $c_{1}$
corresponding to the conserved charge $Q$. The two constants are related
through the expression
\begin{equation}
c_{1}=\sqrt{F_{0}}\kappa^{-1}\sigma^{\frac{1}{2}(3\sigma-1)}(3\sigma
-1)^{\frac{1}{2}(3\sigma-1)}V_{0}^{\frac{1}{2}(1-3\sigma)}.
\end{equation}
By setting \eqref{psidot}, \eqref{apower}, \eqref{FVpower1} and $V(\phi
)=V_{0}\phi^{\kappa}$ into the constraint equation \eqref{con} we can solve
the latter with respect to $\dot{\phi}$. In particular, we obtain:%
\begin{equation}
\label{intermphi1}%
\begin{split}
\dot{\phi}=  &  \pm\bigg[\frac{2 \sigma^{3 \sigma-1} (3 \sigma-1)^{3 \sigma-1}
V_{0}^{1-3 \sigma}}{\kappa^{2}}t^{-6 \sigma} \phi^{\kappa(1 -3 \sigma) }
\left(  2 \phi^{2}-\kappa^{2} \sigma\right)  +\frac{6\sigma^{2}}{t^{2}}%
-2V_{0}\phi^{\kappa}\bigg]^{\frac{1}{2}}.
\end{split}
\end{equation}
Substitution of the $\dot{\phi}$ we derived and of all the rest of the
previous mentioned expressions into equation \eqref{ode1} leads to an
algebraic solution for the scalar field $\phi$:
\begin{equation}
\phi(t)=\left(  \frac{\sigma(3\sigma-1)}{V_{0}}\right)  ^{\frac{1}{\kappa}%
}t^{-\frac{2}{\kappa}}. \label{solphi1}%
\end{equation}
Consequently, the scalar field $\psi$ can be easily obtained now from
\eqref{psidot} with a simple integration. The solution reads:
\begin{equation}%
\begin{split}
\psi(t)=\psi_{0}+  &  2\sigma^{-\frac{3\sigma}{2}}(3\sigma-1)^{\frac{1}%
{2}-\frac{3\sigma}{2}}V_{0}^{-\frac{-3\kappa\sigma+\kappa+4}{2\kappa}%
}t^{3\sigma-1}\\
&  \times\left(  \frac{2(3\sigma-1)^{2/\kappa}\sigma^{\frac{2}{\kappa}%
+\frac{1}{2}}t^{-4/\kappa}}{\kappa(1-3\sigma)+4}+\frac{\kappa\sigma^{3/2}%
V_{0}^{2/\kappa}}{3\sigma-1}\right)  F_{0}^{-\frac{1}{2}},
\end{split}
\label{solpsi1}%
\end{equation}
where $\psi_{0}$ is a constant of integration. One can easily verify that the
$\phi(t)$ and $\psi(t)$ as given by \eqref{solphi1} and \eqref{solpsi1},
together with the scale factor \eqref{apower} satisfy the field equations
\eqref{odes} for the power law potential under consideration and the coupling
function \eqref{FVpower1}. We have thus obtained the solution for the theory,
which admits a power law solution for the scale factor and for a power law
potential $V$ with respect to $\phi$.

\vspace{3pt}

\paragraph{Example 2: Exponential potential.}

We repeat the same procedure only this time we use a potential of the form
$V(\phi)=V_{0}e^{\lambda\phi}$ with $\lambda$ a constant. The compatible (for
a power law $a(t)$) coupling function $F(\phi)$ now becomes
\begin{equation}
F(\phi)=F_{0}e^{\lambda(3\sigma-1)\phi}, \label{FVpower2}%
\end{equation}
where $F_{0}$ is connected to the $c_{1}$ that we saw previously through
\begin{equation}
c_{1}=\sqrt{2}\sqrt{F_{0}}\lambda^{-1}\sigma^{\frac{3\sigma}{2}-\frac{1}{2}%
}(3\sigma-1)^{\frac{3\sigma}{2}-\frac{1}{2}}\sqrt{\lambda^{2}\sigma-2}%
V_{0}^{\frac{1}{2}-\frac{3\sigma}{2}}.
\end{equation}
We notice that unlike the power law potential case, here, an exponential
potential needs also an exponential coupling function to produce a scale
factor of the form \eqref{apower}. As previously done, we substitute
\eqref{psidot}, \eqref{apower}, \eqref{FVpower2} and $V(\phi)=V_{0}%
e^{\lambda\phi}$ into the constraint equation \eqref{con}. From the latter we
obtain the expression for $\dot{\phi}$. Subsequently, we substitute the latter
into equation \eqref{ode1} and we get an algebraic solution for $\phi(t)$
which this time reads
\begin{equation}
\phi(t)=\frac{1}{\lambda}\ln\left(  \frac{\sigma(3\sigma-1)}{t^{2}V_{0}%
}\right)  . \label{solphi2}%
\end{equation}
The direct integration of \eqref{psidot} leads to
\begin{equation}
\psi(t)=\psi_{0}+\frac{\sigma^{\frac{1}{2}-\frac{3\sigma}{2}}(3\sigma
-1)^{-\frac{3\sigma}{2}-\frac{1}{2}}\sqrt{2\lambda^{2}\sigma-4}V_{0}%
^{\frac{3\sigma}{2}-\frac{1}{2}}}{\sqrt{F_{0}}\lambda}t^{3\sigma-1}.
\label{solpsi2}%
\end{equation}
Again, it is easy to check that the above expressions satisfy the field
equations for the choice of potential and the corresponding coupling function
\eqref{FVpower2}. After these two examples we can proceed with the case that
we excluded from the power law solution of the scale factor.

\subsubsection{Case
\texorpdfstring{$\sigma=\frac{1}{3}$}{sigma equal 1 over 3}}

This means that $a(t)=t^{\frac{1}{3}}$, which corresponds to a potential
$V(\phi)=0$. The function $F(\phi)$ remains arbitrary with the solution being
described by the equations
\begin{align}
\label{phisig13}\dot{\psi}  &  = \frac{c_{1}}{t F(\phi)}\\
\dot{\phi}  &  = \pm\frac{1}{t}\left(  \frac{2}{3}- \frac{c_{1}^{2}}{F(\phi
)}\right)  ^{\frac{1}{2}} .
\end{align}
The latter set implies that the following relation exists between the two
scalar fields
\begin{equation}
\psi(\phi) = \psi_{0} \pm\int\!\! \frac{\sqrt{3} c_{1}}{\sqrt{F(\phi)} \sqrt{2
F(\phi)-3 c_{1}^{2}}} d\phi,
\end{equation}
with $\psi_{0}$ being a constant. The function $F(\phi)$ can be chosen
arbitrary and the solution can be obtained through a simple integration since
\eqref{phisig13} yields
\begin{equation}
t = \exp\left(  \pm\int\!\! \left(  \frac{2}{3}- \frac{c_{1}^{2}}{F(\phi
)}\right)  ^{-\frac{1}{2}} \!\! d\phi\right)  .
\end{equation}

\subsection{De Sitter universe}

\label{secds}

In this case we introduce a scale factor of the form $a(t)=e^{\lambda t}$,
where $\lambda$ is a constant. Later in our analysis we shall treat the more
general case $a(t)=e^{\lambda t^{\mu}}$, but here we want to distinguish the
particular solution corresponding to $\mu=1$ due to its importance in
cosmology. The choice $a(t)=e^{\lambda t}$ for the scale factor leads
\eqref{eqV} to become
\begin{equation}
\label{VdS}V(\phi) = 3 \lambda^{2} ,
\end{equation}
Hence, the potential plays the role of a pure cosmological constant.

The constraint equation with the use of \eqref{psidot}, which now reads
\begin{equation}
\label{dotpsidS}\dot{\psi} = \frac{c_{1} e^{-3 \lambda t}}{F(\phi)}%
\end{equation}
yields
\begin{equation}
\label{dotphidS}\dot{\phi} = \pm\frac{c_{1} e^{-3 \lambda t}}{\sqrt{-F(\phi)}}%
\end{equation}
where $F(\phi)$ can be any arbitrary function of $\phi$. As in the special
case of the previous section, we can deduce a relation between $\psi$ and
$\phi$ since
\begin{equation}
\label{dotpsidSsol}\frac{\dot{\psi}}{\dot{\phi}} = \pm\frac{1}{\sqrt{-F(\phi
)}} \Rightarrow\psi= \psi_{0} \pm\int\!\! \frac{1}{\sqrt{-F(\phi)}} d\phi.
\end{equation}
Equation \eqref{dotphidS} can serve to obtain $t$ as a function of $\phi$ and
it results in
\begin{equation}
t = \frac{1}{3\lambda} \ln\left(  \frac{\mp c_{1}}{3 \lambda\int\!\!
\sqrt{-F(\phi)} \, d\phi} \right)  .
\end{equation}

We notice that the previous expressions for $\dot{\psi}$ and $\dot{\phi}$ from
\eqref{dotphidS} and \eqref{dotpsidS} are such so that the sum of the kinetic
terms corresponding to the matter content of the Lagrangian \eqref{LagN} is
zero, i.e.
\begin{equation}
\frac{a^{3} F(\phi) \dot{\psi}^{2}}{2 N}+\frac{a^{3} \dot{\phi}^{2}}{2 N}
=\frac{1}{2} e^{3 \lambda t} \left(  F(\phi) \dot{\psi}^{2}+ \dot{\phi}%
^{2}\right)  = 0.
\end{equation}
Thus, it is necessary for one of the kinetic terms to change its sign. This
was expected from the moment that we noticed that the choice $a(t)=e^{\lambda
t}$ led to a potential that is a constant. We already know that such a
space-time is a pure cosmological constant solution, hence any other
contributions should in principle cancel each other in the Lagrangian. The
only other possibility in the theory under consideration would the trivial
solution of $\phi$ and $\psi$ being constants.

In the particular case where $F(\phi)=$const. we have the typical quintessense
- phantom pair that is usually encountered in quintom scenarios. In their
generality quintom cosmological models are constructed so that the boundary
$w=-1$ is crossed \cite{qui01,qui02}, where $w$ is the usual equation of state
parameter. In this paradigm we are trivially on this boundary with the kinetic
energies being cancelling each other for every $t$.

\subsection{Asymptotically de Sitter space-time}

In this section we examine a couple of cases where the space-time can be
considered as being asymptotically de Sitter in the sense that, locally, the
metric becomes that of a positive cosmological constant solution in the limit
$t\rightarrow\pm\infty$.

\subsubsection{Space-time without singularity}

We adopt for the scale factor the choice $a(t)=\left(  \cosh(t)\right)
^{\alpha}$. The resulting spacetime has its curvature scalars regular both at
$t\rightarrow0$ and $t\rightarrow\pm\infty$. We can distinguish two cases
depending on whether the constant $\alpha$ takes the value $\frac{1}{3}$ or
not. We start from the generic case, so we assume $a\neq\frac{1}{3}$. Under
this assumption \eqref{eqV} becomes
\begin{equation}
V(\phi)=\alpha+\alpha(3\alpha-1)\tanh^{2}(t), \label{Vtcosh}%
\end{equation}
which can be inverted to
\begin{equation}
t=\pm\mathrm{arctanh}\left(  \frac{\sqrt{V(\phi)-\alpha}}{\sqrt{\alpha
(3\alpha-1)}}\right)  . \label{tVcosh}%
\end{equation}
For large values of $t$, the potential (\ref{Vtcosh}) becomes $V\left(
\phi\right)  \simeq3\alpha^{2}$ from where the de Sitter solution $a\left(
t\right)  \simeq e^{\alpha t}$ is recovered.

If we take the total derivative with respect to time, equation \eqref{Vtcosh}
leads to
\begin{equation}
\dot{\phi}=\frac{2\alpha(3\alpha-1)\tanh(t)}{\cosh(t)^{2} V^{\prime}(\phi)}.
\label{phiV2}%
\end{equation}
Substitution of \eqref{tVcosh}, \eqref{phiV2} and \eqref{psidot} into the
constraint equation \eqref{con} provides us with a relation between the two
functions of $\phi$
\begin{equation}
F(\phi)=\frac{c_{1}^{2}\alpha^{1-3\alpha}(1-3\alpha)^{1-3\alpha}\left(
V(\phi)-3\alpha^{2}\right)  ^{3\alpha-1}V^{\prime}(\phi)^{2}}{2\left(
6\alpha^{3}-\alpha V^{\prime}(\phi)^{2}-2(3\alpha+1)\alpha V(\phi
)+2V(\phi)^{2}\right)  }. \label{FVnons}%
\end{equation}

For completeness, we need to also consider the case $\alpha=\frac{1}{3}$. When
this value for the power of $\cosh(t)$ is adopted, relation \eqref{eqV}
results in a constant potential $V(\phi)=\frac{1}{3}$. The constraint equation
\eqref{con} and the first integral $Q$ yield:
\begin{align}
\label{phicosh}\dot{\psi}  &  = \frac{c_{1} }{F(\phi)\cosh(t)}\\
\dot{\phi}  &  = \pm\frac{1}{\cosh(t)}\sqrt{-\frac{c_{1}^{2}}{F(\phi)}%
-\frac{2}{3}}.
\end{align}
From the above we extract easily the relation
\begin{equation}
\frac{\dot{\psi}}{\dot{\phi}} = \pm\frac{c_{1}}{\sqrt{-\frac{1}{3} F(\phi)
\left(  3 c_{1}^{2}+2 F(\phi)\right)  }} \Rightarrow\psi= \psi_{0} \pm
\int\!\!\frac{c_{1}}{\sqrt{-\frac{1}{3} F(\phi) \left(  3 c_{1}^{2}+2
F(\phi)\right)  }} d\phi.
\end{equation}
What is more, equation \eqref{phicosh} is separable and we can easily derive
$t$ in terms of $\phi$ for any (admissible) function $F(\phi)$,
\begin{equation}
t = 2 \tanh^{-1}\left[  \tan\left(  \pm\frac{1}{2} \int\!\! \frac{1}%
{\sqrt{-\frac{c_{1}^{2}}{F(\phi)}-\frac{2}{3}}} d\phi\right)  \right]  .
\end{equation}

\subsubsection{Space-time with a curvature singularity}

If we consider a scale factor of the form $a(t) = (\sinh(t))^{\alpha}$ we
obtain a space-time that possesses a curvature singularity at $t\rightarrow0$,
as is evident by the Ricci scalar that reads
\begin{equation}
R=\frac{6 \alpha(\alpha+\alpha\cosh(2 t)-1)}{\sinh^{2}(t)} .
\end{equation}
Once more we have to distinguish cases depending on the parameter $\alpha$. We
start by considering $\alpha\neq\frac{1}{3}$. The same analysis that we
followed previously results in the potential
\begin{equation}
\label{Vtsinh}V(\phi) = \alpha+\alpha(3 \alpha-1) \coth^{2}(t),
\end{equation}
which in its turn leads to
\begin{equation}
\label{tVsinh}t = \pm\mathrm{arccoth}\left(  \frac{\sqrt{V(\phi)-\alpha}%
}{\sqrt{\alpha( 3\alpha-1) }}\right)  .
\end{equation}
and
\begin{equation}
\label{phiV2b}\dot{\phi} = \frac{2 \alpha(1-3 \alpha) \coth(t) }{\sinh(t)^{2}
V^{\prime}(\phi)}.
\end{equation}

With the help of the above the constraint equation \eqref{con} gives rise to
\begin{equation}
F(\phi)=-\frac{c_{1}^{2}\alpha^{1-3\alpha}(3\alpha-1)^{1-3\alpha}\left(
V(\phi)-3\alpha^{2}\right)  ^{3\alpha-1}V^{\prime}(\phi)^{2}}{2\left(
6\alpha^{3}-\alpha V^{\prime}(\phi)^{2}-2(3\alpha+1)\alpha V(\phi
)+2V(\phi)^{2}\right)  }, \label{FVsing}%
\end{equation}
which for integer values of $\alpha$ can be either equal to \eqref{FVnons} or
opposite of it. In particular, if $\alpha$ is an odd number \eqref{FVsing} is
the opposite of \eqref{FVnons}. On the other hand, if it is even, then the two
expressions yield the same result. An interesting value for the parameter
$\alpha$ is that of $\alpha=\frac{2}{3}$, where the model mimics the $\Lambda$-cosmology.

Finally, we conclude this section by considering the $\alpha=\frac{1}{3}$
case. Once more the potential assumes the constant value $V(\phi)=\frac{1}{3}%
$, while for the two scalar fields we have
\begin{align}
\dot{\psi}  &  =\frac{c_{1}}{F(\phi)\sinh(t)}\\
\dot{\phi}  &  =\pm\frac{1}{\sinh(t)}\sqrt{\frac{2}{3}-\frac{c_{1}^{2}}%
{F(\phi)}}. \label{singphid}%
\end{align}
By combining the last two relations we get
\begin{equation}
\frac{\dot{\psi}}{\dot{\phi}}=\pm\sqrt{\frac{3c_{1}^{2}}{F(\phi)\left(
2F(\phi)-3c_{1}^{2}\right)  }}\Rightarrow\psi=\psi_{0}\pm\int\!\!\sqrt
{\frac{3c_{1}^{2}}{F(\phi)\left(  2F(\phi)- 3c_{1}^{2}\right)  }}d\phi,
\end{equation}
while \eqref{singphid} implies
\begin{equation}
t = 2 \tanh^{-1} \left[  \exp\left(  \pm\int\!\! \frac{1}{\sqrt{\frac{2}%
{3}-\frac{c_{1}^{2}}{F(\phi)}}}d\phi\right)  \right]
\end{equation}

\subsection{The
\texorpdfstring{$a(t)=e^{\lambda t^\mu}$}{a(t)=exp (lambda t raised in mu)}
case}

In this section we examine the effect of a scale factor of the form
$a(t)=e^{\lambda t^{\mu}}$ with $\mu\neq1$ (remember that the $\mu=1$ case was
studied separately in section \ref{secds}). For $\lambda<0$ the scale factor
can be said that follows a generalized normal distribution (the particular
constant factor needed to multiply the exponent can be provided by scaling
appropriately the spatial distance $r$ in the metric). With this specific
example we want to demonstrate that the method we propose can lead to analytic
solution even under the assumption of more complicated scale factors. In what
regards the singularities of the space-time, it can be observed from the
scalar curvature
\begin{equation}
R = 6 \lambda\mu t^{\mu-2} \left(  \mu-1 +2 \lambda\mu t^{\mu}\right)  ,
\end{equation}
that the latter diverges at $t\rightarrow\infty$ when $\mu>1$ and at $t=0$
when $\mu\in(-\infty,2)-\{1\}$. With this choice of a scale factor equation
\eqref{eqV} is written as
\begin{equation}
\label{Vtcomb1}\frac{3 \lambda^{2} \mu^{2} }{V(\phi)}t^{2 \mu} + \frac
{\lambda(\mu-1) \mu}{V(\phi)} t^{\mu} = t^{2} .
\end{equation}
We need to distinguish two cases depending on whether $\mu=2$ or not,
something that is going to become obvious in the subsequent analysis.

\subsubsection{Case \texorpdfstring{$\mu\neq 1, 2$}{mu not equal to 1 or 2}}

In this setting it is interesting to observe that the above equation can be
solved algebraically with respect to $t$ even for an arbitrary power $\mu$.
The solution can be given in terms of nested radicals in a manner similar that
the solution of the more general equation of the form \cite{Bagis}
\[
\alpha t^{2\mu} + \beta t^{\mu}+\gamma= t^{\nu}.
\]
In our case we have $\nu=2$ and $\gamma=0$. This value for $\gamma$ leads us
to a small modification of the process of deriving the solution in comparison
to what is presented in \cite{Bagis}. We first notice that \eqref{Vtcomb1} can
be written as
\begin{equation}
\label{intermidt}t^{\mu}= -\frac{\mu-1}{3 \lambda\mu}+\frac{V(\phi)}{3
\lambda^{2} \mu^{2}}t^{2-\mu} \Rightarrow t= \sqrt[\mu]{-\frac{\mu-1}{3
\lambda\mu}+\frac{V(\phi)}{3 \lambda^{2} \mu^{2}}t^{2-\mu}}.
\end{equation}
We raise this last relation to the power $2-\mu$ (we remind that we assume
$\mu\neq2$) and we obtain
\begin{equation}
\label{t2mu}t^{2-\mu}= \sqrt[\mu/(2-\mu)]{-\frac{\mu-1}{3 \lambda\mu}%
+\frac{V(\phi)}{3 \lambda^{2} \mu^{2}}t^{2-\mu}} .
\end{equation}
This means that the $t$ in \eqref{intermidt} may now be given in terms of an
infinite number of radicals through successive substitutions of \eqref{t2mu}.
In particular we get:
\begin{equation}
\label{finaltV}t = \sqrt[\mu]{-\frac{\mu-1}{3 \lambda\mu}+\frac{V(\phi)}{3
\lambda^{2} \mu^{2}} \sqrt[\mu/(2-\mu)]{-\frac{\mu-1}{3 \lambda\mu}%
+\frac{V(\phi)}{3 \lambda^{2} \mu^{2}} \sqrt[\mu/(2-\mu)]{-\frac{\mu-1}{3
\lambda\mu}+\frac{V(\phi)}{3 \lambda^{2} \mu^{2}}\ldots}}}.
\end{equation}
By obtaining the expression for $\dot{\phi}$ from the time derivative of
\eqref{Vtcomb1} and by using equation \eqref{psidot} into the constraint
\eqref{con} we are led to
\begin{equation}
\label{FVfinal}F(\phi) = -\frac{c_{1}^{2} e^{-6 \lambda t^{2}} V^{\prime}%
(\phi)^{2}}{\lambda^{2} (\mu-1)^{2} \mu^{2} t^{2 (\mu-3)} \left(  6 \lambda\mu
t^{\mu}+\mu-2\right)  ^{2}+ \left(  2 V(\phi)-24 \lambda^{2} t^{2}\right)
V^{\prime}(\phi)^{2} },
\end{equation}
where $t$ is understood to be given in terms of $V(\phi)$ from
\eqref{finaltV}. So $F(\phi)$ and $V(\phi)$ need to also be related with an
expression involving nested radicals in order to acquire a solution of the
form $a(t)=e^{\lambda t^{\mu}}$ with $\mu\neq2$. We now proceed to complete
this section with the study of the system when $\mu=2$.

\subsubsection{The \texorpdfstring{$\mu=2$}{mu equal to 2} case}

This situation corresponds to a normal distribution scale factor whenever
$\lambda<0$. However, in what follows we make no particular assumption over
the sign of $\lambda$. By setting $\mu=2$ in \eqref{Vtcomb1} we immediately
see that it results in
\begin{equation}
\label{ttoVmu2}t = \pm\frac{\sqrt{V(\phi)-2 \lambda}}{2 \sqrt{3} \lambda}.
\end{equation}
At the same time from its time derivative we get
\begin{equation}
\label{phiVmu2}\dot{\phi} = \frac{24 \lambda^{2} t}{V^{\prime}(\phi)}.
\end{equation}
By following the exact same procedure as in the previous cases, the above
relations help us arrive at
\begin{equation}
\label{FVmu2}F(\phi) = \frac{c_{1} ^{2} e^{1-\frac{V(\phi)}{2 \lambda}}
V^{\prime2}(\phi)}{96 \lambda^{3}-4 \lambda V^{\prime2}-48 \lambda^{2}
V(\phi)} .
\end{equation}
We study a quick example that demonstrates the application of relation \eqref{FVmu2}.

\vspace{3pt}

\paragraph{Example: Power Law potential.}

We choose to set $V(\phi) = V_{0} \phi^{\kappa}+ \Lambda$, where $\Lambda$ is
a constant. According to \eqref{FVmu2} the compatible $F(\phi)$ function, for
the scale factor we want, is
\begin{equation}
\label{FVmuex2}F(\phi) = \frac{c_{1}^{2} \kappa^{2} V_{0}^{2} e^{1-\frac
{\Lambda+ V_{0} \phi^{\kappa}}{2 \lambda}}\phi^{2( \kappa-1)} }{96 \lambda
^{3}-4 \kappa^{2} \lambda V_{0}^{2} \phi^{2( \kappa-1)}-48 \lambda^{2} \left(
\Lambda+ V_{0} \phi^{\kappa}\right)  }.
\end{equation}
As in the process of deriving the solutions in the previous examples, we need
to use \eqref{psidot} and \eqref{FVmuex2} into the constraint equation and
solve algebraically with respect to $\dot{\phi}$. Substitution of this result
into equation \eqref{ode1} for the given form of the potential $V(\phi)$
brings about the solution
\begin{equation}
\phi(t) = \left(  \frac{2 \lambda-\Lambda+12 \lambda^{2} t^{2}}{V_{0}}\right)
^{1/\kappa} .
\end{equation}
The above result leads \eqref{psidot} to yield
\begin{equation}
\psi(t) = \psi_{0} - \int\!\! \frac{4 \lambda e^{3 \lambda t^{2}}}{c_{1}%
}\left[  \frac{ 144 \lambda^{3} t^{2} V_{0}^{-2/\kappa}}{\kappa^{2}}\left(  2
\lambda\left(  6 \lambda t^{2}+1\right)  -\Lambda\right)  ^{\frac{2}{\kappa
}-2}+1 \right]  dt.
\end{equation}
Thus, we derived the solution that allows for $a(t)= e^{\lambda t^{2}}$ when a
power law potential is present.

We summarise all the results obtained in this section, involving the effect
that the various scale factors have in the relation between $F(\phi)$ and
$V(\phi)$, in table \eqref{TableKey}.%

\begin{table}[tbp] \centering
\caption{Relation between functions $F(\phi)$ and $V(\phi)$ for specific functional forms of the scale factor $a(t)$}%
\begin{tabular}
[c]{ccc}\hline\hline
$a(t)$ & $F-V$ relation & $F$ arbitrary\\\hline
$t^{\sigma}$ & eq. \eqref{FVpower} if $\sigma\neq\frac{1}{3}$ & $V(\phi)=0$ if
$\sigma=\frac{1}{3}$\\
$\cosh^{\alpha}(t)$ & eq. \eqref{FVnons} if $\alpha\neq\frac{1}{3}$ &
$V(\phi)=\frac{1}{3}$ if $\alpha=\frac{1}{3}$\\
$\sinh^{\alpha}(t)$ & eq. \eqref{FVsing} if $\alpha\neq\frac{1}{3}$ &
$V(\phi)=\frac{1}{3}$ if $\alpha=\frac{1}{3}$\\
$e^{\lambda t^{\mu}}$ & \eqref{FVfinal} if $\mu\neq1,2$ or \eqref{FVmu2} if
$\mu=2$ & $V(\phi)=3\lambda^{2}$ if $\mu=1$\\\hline\hline
\end{tabular}
\label{TableKey}%
\end{table}%

\section{Stability of the scaling solutions for the exponential potential}

\label{sec4}

We now proceed by studying the stability of the scaling solutions in the case
where $V\left(  \phi\right)  $ and $F\left(  \phi\right)  $ are exponential
functions, i.e. $V\left(  \phi\right)  =V_{0}e^{\lambda\phi}$ and $F\left(
\phi\right)  =F_{0}e^{\mu\phi}$. In this consideration the scalar field theory
is reduced to the Chiral cosmology.

We introduce the new dimensionless variables \cite{copeland} (where for the
lapse function we assume $N=1$).%
\begin{equation}
x=\frac{\dot{\phi}}{\sqrt{6}H}~,~y=\frac{\sqrt{V\left(  \phi\right)  }}%
{\sqrt{3}H}~,~z=\frac{\sqrt{F\left(  \phi\right)  }\dot{\psi}}{\sqrt{6}H},
\label{p1}%
\end{equation}
where $H=\frac{\dot{a}}{a}$. The field equations are written as the following
system%
\begin{equation}
x^{2}+y^{2}+z^{2}-1=0, \label{p2}%
\end{equation}%
\begin{equation}
\frac{dx}{d\tau}=\frac{3}{2}x\left(  x^{2}+z^{2}-y^{2}-1\right)  +\frac
{\sqrt{6}}{2}\left(  \mu z^{2}-\lambda y^{2}\right)  , \label{p3}%
\end{equation}%
\begin{equation}
\frac{dy}{d\tau}=\frac{3}{2}y\left(  x^{2}+z^{2}-y^{2}+1\right)  +\frac
{\sqrt{6}}{2}xy\lambda, \label{p4}%
\end{equation}%
\begin{equation}
\frac{dz}{d\tau}=\frac{3}{2}z\left(  x^{2}+z^{2}-y^{2}-1\right)  -\frac
{\sqrt{6}}{2}xz\mu. \label{p5}%
\end{equation}
in which the new independent variable $\tau$ is defined as $\tau=\ln\left(
a\left(  t\right)  \right)  $.

From (\ref{p2}) it is clear that the variables $\left\{  x,y,z\right\}  $
evolve on a three-dimensional sphere with $y\geq0$. Moreover, every critical
point of the latter system corresponds to a phase where the scale factor
$a\left(  t\right)  $ is given by the expressions $a\left(  t\right)
=t^{\frac{2}{3\left(  1+w_{tot}\right)  }}$ when $w_{tot}\neq-1$ or $a\left(
t\right)  =e^{H_{0}t}$ when $w_{tot}=-1$ in which $w_{tot}$ is the equation of
state parameter for the effective fluid defined as%
\begin{equation}
w_{tot}\left(  x,y,z\right)  =x^{2}+z^{2}-y^{2}. \label{P6}%
\end{equation}

The critical points of the system (\ref{p2})-(\ref{p5}) are determined to be%
\begin{equation}
P_{A}^{\left(  \pm\right)  }=\left(  \pm1,0,0\right)  ~,~P_{B}=\left(
-\frac{\lambda}{\sqrt{6}},\frac{\sqrt{6-\lambda^{2}}}{6},0\right)  ,
\end{equation}%
\begin{equation}
P_{C}^{\left(  \pm\right)  }=\left(  -\frac{\sqrt{6}}{\lambda+\mu},\sqrt
{\frac{\mu}{\lambda+\mu}},\pm\frac{\sqrt{\lambda^{2}+\lambda\mu-6}}%
{\lambda+\mu}\right)  .
\end{equation}

Points $P_{A}^{\left(  \pm\right)  }$ and $P_{B}$ correspond to the scaling
solution where the field $\psi$ does not contribute to the evolution of the
universe. Points $P_{A}^{\left(  \pm\right)  }$ and $P_{B}$ are defined in the
surface $z=0$ and are given in \cite{copeland}. Points $P_{C}^{\left(
\pm\right)  }$ are the new points that emerge from our analysis. We continue
with the discussion of the physical quantities of the critical points as also
with their stability. In order to study the latter, we reduce the dynamical
system into a two-dimensional system given by the equations (\ref{p3}) and
(\ref{p4}), where from (\ref{p2}) it follows $z^{2}=1-x^{2}-y^{2}$.

As far as the points $P_{A}^{\left(  \pm\right)  }$ are concerned, the
universe is dominated by the kinetic term of the scalar field $\phi$, i.e.
$w_{tot}=1$. \ The eigenvalues of the linearized system are determined to be
\begin{equation}
e_{1}\left(  P_{A}^{\left(  \pm\right)  }\right)  =3~,~e_{2}\left(
P_{A}^{\left(  \pm\right)  }\right)  =3\pm\frac{\sqrt{6}}{2}\lambda
\end{equation}
from where we infer that the points are always unstable because $e_{1}\left(
P_{A}^{\left(  \pm\right)  }\right)  $ is always positive.

Point $P_{B}$ exists only when $\lambda^{2}<6$ and describes a scaling
solution where $w_{tot}=-1+\frac{\lambda^{2}}{3}$. The eigenvalue of the
linearized system are determined to be%
\begin{equation}
e_{1}\left(  P_{B}\right)  =-3+\lambda^{2}~,~e_{2}\left(  P_{B}\right)
=-3+\frac{\lambda^{2}}{2},
\end{equation}
from where we conclude that the solution at point $P_{B}$ is stable when
$\lambda^{2}<3$. The universe is accelerated when $\lambda^{2}<2$.

\begin{figure}[ptb]
\includegraphics[width=0.4\textwidth]{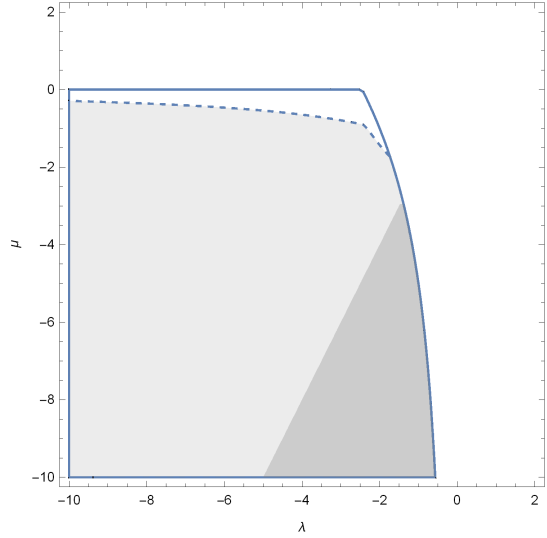}\caption{Region plot in the
space of variables $\left\{  \lambda,\mu\right\}  $ for the eigenvalues
$e_{1}\left(  P_{C}\right)  $, $e_{2}\left(  P_{C}\right)  $. With the solid
line is the boundary where the points exist. Dashed line describe the boundary
where the points are attractors. The ope gray area \ corresponds to the case
where $w_{tot}>-\frac{1}{3}$, while the dark gray area describe the values of
the parameters $\left\{  \lambda,\mu\right\}  $ in which $w_{tot}<-\frac{1}%
{3}$. }%
\label{fig1}%
\end{figure}

In what regards the new points $P_{C}$, attained from our analysis, they
describe a scaling solution with $w_{tot}=-1+\frac{2\lambda}{\lambda+\mu}$
which corresponds to an accelerated universe when $\left\{  \mu<0,\frac{\mu
}{2}<\lambda<\left\vert \mu\right\vert \right\}  $ or $\left\{  \mu
>0,~-\mu<\lambda<\frac{\mu}{2}\right\}  $. This is the solution we previously
derived. It is important to mention that the points exist when $\lambda
\neq-\mu$ and $\left\{  \mu<0,~\lambda\leq-\sqrt{6}\right\}  $ and $\left\{
-\sqrt{6}<\lambda<0,~\mu<\frac{6-\lambda^{2}}{\lambda}\right\}  $ from where
we infer that $w_{tot}<-\frac{1}{3}$ when $\left\{  \lambda\leq-\sqrt{3}%
,\mu<\lambda\right\}  $ or $\left\{  -\sqrt{3}<\lambda<0,~~\mu<\frac
{6-\lambda^{2}}{\lambda}\right\}  $. \ As far as the stability is concerned,
because of the nonlinearity dependence of the eigenvalues $e_{1}\left(
P_{C}\right)  ,~e_{2}\left(  P_{C}\right)  $ on the free parameters $\mu,$ and
$\lambda$, we solve numerically the conditions where $e_{1}\left(
P_{C}\right)  <0,~e_{2}\left(  P_{C}\right)  <0$, and derive when the points
are attractors. The region in the space $\left\{  \lambda,\mu\right\}  $ where
the points are attractors are given in Fig. \ref{fig1}.

\begin{figure}[ptb]
\includegraphics[width=0.4\textwidth]{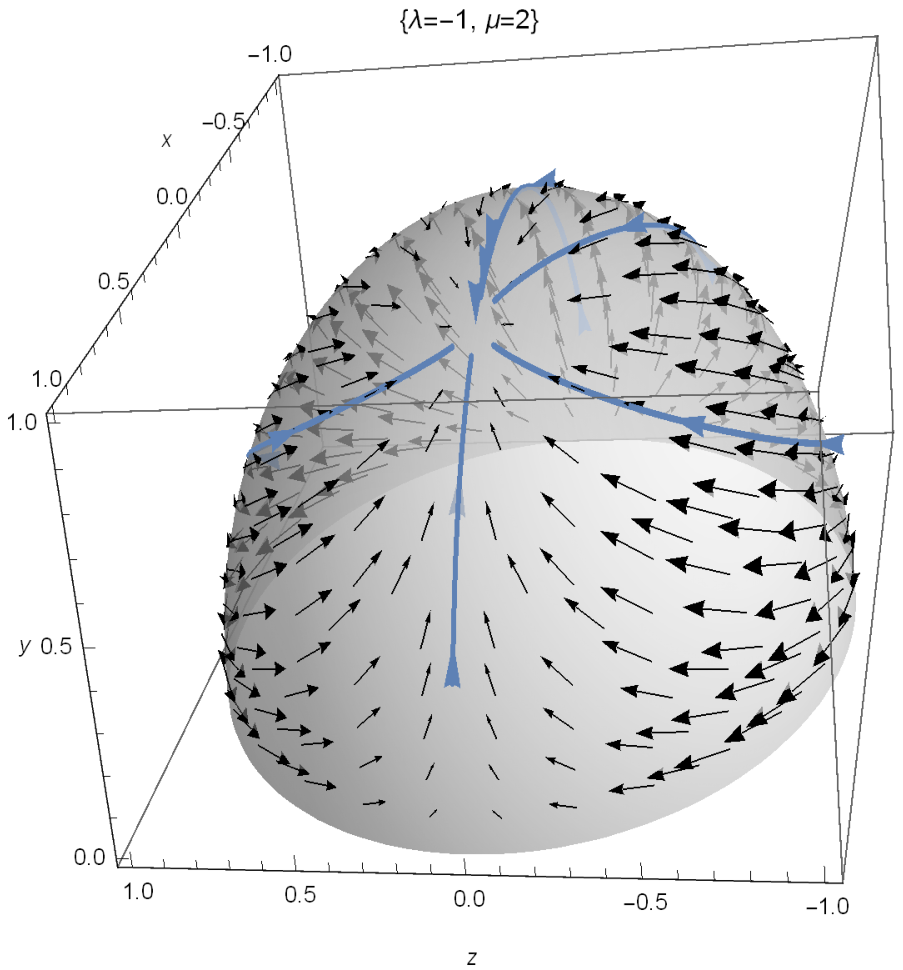}\includegraphics[width=0.5\textwidth]{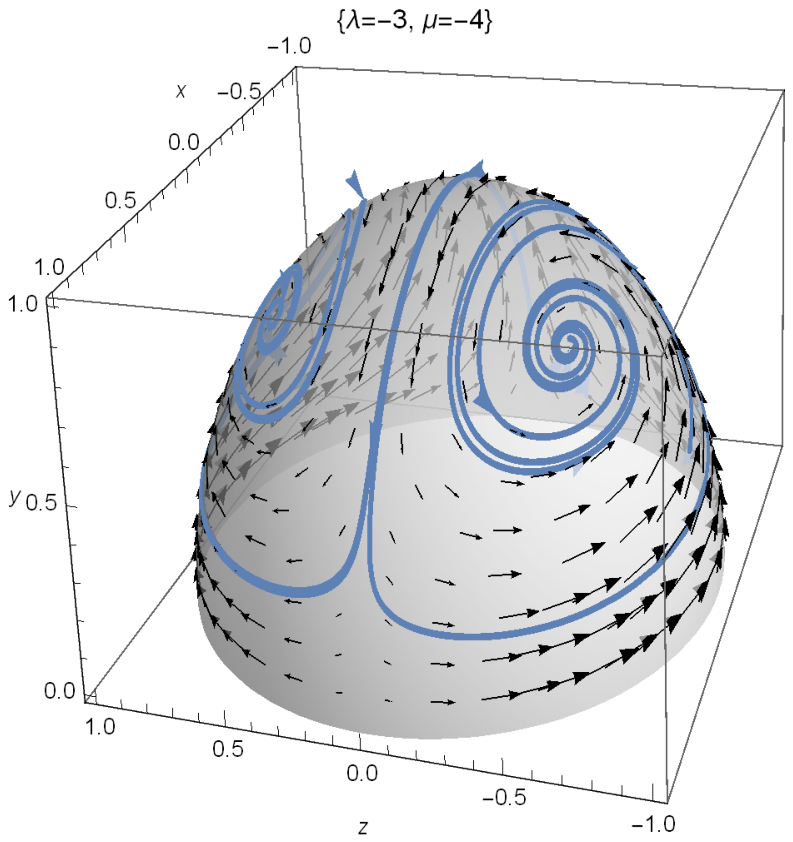}\caption{Phase
space diagram for the dynamical system (\ref{p3})-(\ref{p5}) for two different
set of the free parameters $\lambda$ and$~\mu$. Left figure is for $\left(
\lambda,\mu\right)  =\left(  -1,2\right)  $ where point $P_{B}$ is the unique
attractor while right figure is for $\left(  \lambda,\mu\right)  =\left(
-3,-4\right)  $ and points $P_{C}^{\pm}$ are the attractors of the dynamical
system. }%
\label{fig2}%
\end{figure}

In Fig. \ref{fig2} we present the phase space diagram in the variables
$\left\{  x,y,z\right\}  $ for the dynamical system of our consideration for
two set of the values $\lambda$ and $\mu.$ More specifically, the plots are
\ for the set $\left(  \lambda,\mu\right)  =\left(  -1,2\right)  $ where the
only stable point is $P_{B}$ and for $\left(  \lambda,\mu\right)  =\left(
-3,-4\right)  $ where points $P_{C}^{\left(  \pm\right)  }$ are the two
attractors of the dynamical system.

We summarize the results of the critical point analysis in Table \ref{taa}.%

\begin{table}[tbp] \centering
\caption{Critical points of the dynamical system}%
\begin{tabular}
[c]{ccccc}\hline\hline
\textbf{Point} & $\left(  \mathbf{x,y,z}\right)  $ & \textbf{Existence} &
\textbf{Stability} & \textbf{Acceleration}\\\hline
$P_{A}^{\left(  \pm\right)  }$ & $\left(  \pm1,0,0\right)  $ & Always &
Unstable & No\\
$P_{B}$ & $\left(  -\frac{\lambda}{\sqrt{6}},\frac{\sqrt{6-\lambda^{2}}}%
{6},0\right)  $ & $\lambda^{2}<6$ & Stable: $\lambda^{2}<3$ & $\lambda^{2}%
<2$\\
$P_{C}^{\left(  \pm\right)  }$ & $\left(  -\frac{\sqrt{6}}{\lambda+\mu}%
,\sqrt{\frac{\mu}{\lambda+\mu}},\pm\frac{\sqrt{\lambda^{2}+\lambda\mu-6}%
}{\lambda+\mu}\right)  $ &
\begin{tabular}
[c]{c}%
$\lambda\neq-\mu$\\
$~\left\{  \mu<0,~\lambda\leq-\sqrt{6}\right\}  $\\
$\left\{  -\sqrt{6}<\lambda<0,~\mu<\frac{6-\lambda^{2}}{\lambda}\right\}  $%
\end{tabular}
& See Fig. \ref{fig1} & See Fig. \ref{fig2}\\\hline\hline
\end{tabular}
\label{taa}%
\end{table}%

Let us now discuss in some detail the new critical point $P_{C}$: The
cosmological fluid source at this critical point has an equation of state
parameter $w_{tot}=-1+\frac{2\lambda}{\lambda+\mu}$, therefore we can easily
calculate that
\begin{equation}
a\left(  t\right)  =t^{\frac{1}{3}\left(  1+\frac{\mu}{\lambda}\right)
}~,~H\left(  t\right)  =\frac{\lambda+\mu}{3\lambda}\frac{1}{t}.
\end{equation}
Moreover, from (\ref{p3})-(\ref{p5}) and (\ref{p1}) it follows that at the
critical point
\begin{equation}
\frac{dx}{d\tau}=0~,~\frac{dy}{d\tau}=0~,~\frac{dz}{d\tau}=0
\end{equation}
that is
\[
\dot{\phi}=x_{0}\sqrt{6}H~,~V_{0}e^{\lambda\phi}=\left(  y_{0}\right)
^{2}H^{2}~,~\dot{\psi}=\sqrt{6}z_{0}e^{-\frac{1}{2}\mu\phi}H
\]
which implies
\begin{equation}
\dot{\phi} =\sqrt{6}x_{0}\frac{\lambda+\mu}{3\lambda}\frac{1}{t}%
~~,~e^{\lambda\phi}=\left(  y_{0}\frac{\lambda+\mu}{3\lambda}\right)
^{2}\frac{1}{t^{2}}~,~\dot{\psi}=\sqrt{6}z_{0}\frac{\lambda+\mu}{3\lambda
}t^{-\frac{\sqrt{6}x_{0}}{6}\frac{\mu}{\lambda}\left(  \lambda+\mu\right)  -1}%
\end{equation}
we finally find
\[
\phi\left(  t\right)  -\phi_{0}=\sqrt{\frac{2}{3}}\frac{x_{0}\left(
\lambda+\mu\right)  }{\lambda}\ln t~~,~\psi\left(  t\right)  -\psi_{0}%
=\sqrt{\frac{2}{3}}\frac{\lambda+\mu}{\mu}z_{0}e^{-\frac{\mu}{2}\phi_{0}%
}t^{\frac{\mu}{\lambda}}%
\]
with the constants $x_{0}$ and $y_{0}$ being constrained as $x_{0}%
=-\frac{\sqrt{6}}{\lambda+\mu},~\left(  y_{0}\right)  ^{2}=\frac{9\lambda
^{2}V_{0}}{\left(  \lambda+\mu\right)  ^{2}}e^{\lambda\phi_{0}}$. This is
precisely the exact solution (\ref{solphi2}), (\ref{solpsi2}) determined before.

As we discussed above, when $\frac{2\lambda}{\lambda+\mu}<\frac{2}%
{3}$ the cosmological solution at point $P_{C}$  describes an
accelerated universe, while the attractor where the two scalar fields survive
has been used to describe the inflationary era. When the field space is a
hyperbolic plane \cite{brown}, the model is called hyperinflation. There are
various differences between hyperinflation and the slow-roll inflation
\cite{LiddleBarrow}, for an elaborate discussion see \cite{brown}. Remark that
point $P_{B}$  describe the slow-roll inflation era since only field $\phi$ contribute to the cosmological evolution. On the other hand, $P_C$ corresponds to the two field case and particularly to a solution that emerges for a coupling function $F(\phi) = F_0 e^{\mu \phi}$, which makes the two dimensional field space hyperbolic. 

Recently in \cite{ref1903} the dynamics of the later cosmological model has
been studied by using a different set of variables. Our analysis can be
directly compared with that of \cite{ref1903} and shows that the use of a
different set of variables leaves the dynamics invariant. The critical point
$P_{C}$ corresponds to the point $\left(  x,y\right)  _{hyper}$ of
\cite{ref1903} (corresponding to the $x$, $z$ of our analysis) where the
spiralling behaviour, as described in \cite{ref1903}, is given by eigenvalues
with nonzero imaginary part; such a behaviour is presented in Fig. \ref{fig2}.

\section{Conclusion}

\label{sec5}

We considered a two-scalar field cosmological model with a mixed kinetic term.
More specifically, we assume that the matter source of the Einstein's General
Relativity is described by two scalar field which they have an interacting
term in the kinetic part of the scalar field Lagrangian. For this model, we
investigate the existence of some exact solutions for the scale factor which
describe different phases in the evolution of the universe.

In particular, we proved the existence of power-law solutions, de Sitter
universe, asymptotic de Sitter universe with or without initial singularity
and for a scale factor that depends on the exponential function of arbitrary
powers of the time variable. For each specific solution we found the
constraining condition between the two unknown functions of the theory so that
it would be admissible. These unknown functions are the scalar field potential
$V(\phi)$ and the one that provides the interaction in kinetic term $F(\phi)$.
A special case of the model which we considered is the Chiral cosmology, which
can be seen as a unified dark energy and dark matter model. We also
demonstrated the validity of our result by providing several examples in which
we choose the potential function $V(\phi)$ and derive the compatible $F(\phi)$
for which the space-time of our choice is admissible.

For the Chiral cosmological model and for the exponential potential we apply
the critical point analysis in order to determine the stability of the scaling
solutions. We found that the dynamical system admits five critical points
where the three of them correspond to the quintessence model of General
Relativity. The two new critical points describe scaling solutions and by our
analysis we were able to study the stability of the exact solutions.

Apart from the approach that we followed here in which,
when certain conditions are set over the scale factor,
the potential is related to the cosmological time through equation %
\eqref{eqV}, a different procedure can also be applied: It is known that, in the case of a single scalar field, when the latter is
utilized as an effective time variable, you can express the potential as a
function of quantities containing only $\phi$ \cite{Salopek,LiddleBarrow}.
This is made possible with the use of the equations of motion and is called
the Hamilton-Jacobi method. A similar methodology has been put in use recently
in \cite{Achucarro1,Achucarro2} for the case of multiple scalar fields where
the potential is expressed in terms of the fields and their first derivatives.
Especially in \cite{Achucarro2} - and for a two scalar field system - the
adoption of the a shift-symmetric orbital inflation condition leads to
expressing the potential as a pure function of the two fields while at the
same time providing with an exact solution. It is interesting to note that
also in this approach one obtains a relation between the potential and the
kinetic coupling function, although under a different context since the
potential is assumed to depend on both fields.

In a future work we plan to investigate the case where the two scalar fields
are interacting also in the potential term, as also when they are nonminimally
coupled with the gravity \cite{re4}.

\end{document}